\documentclass{aa}  

\usepackage{graphicx}
\usepackage{multirow}
\usepackage{txfonts}
\usepackage[colorlinks,allcolors=blue]{hyperref}

\newcommand\figsize{0.45}
\newcommand{\rahour}{\hbox{\ensuremath{^{\rm h}}}}
\newcommand{\ramin}{\hbox{\ensuremath{^{\rm m}}}}
\newcommand{\xmm}{{\it XMM-Newton}\xspace}
\newcommand{\suzaku}{{\it Suzaku}\xspace}
\newcommand{\swift}{{\it Swift}\xspace}
\newcommand{\gaia}{{\it Gaia}\xspace}

\begin{document}

   \title{An intermediate polar candidate toward the Galactic plane}

    \author{S. Mondal\inst{1},
          G. Ponti\inst{1,2},
          F. Haberl\inst{2},
          K. Anastasopoulou \inst{1},
          S. Campana\inst{1},
          K. Mori\inst{3},
          C. J. Hailey\inst{3}, and 
          N. Rea\inst{4,5}
          }

   \institute{$^1$INAF – Osservatorio Astronomico di Brera, Via E. Bianchi 46, 23807 Merate (LC), Italy
             \email{samaresh.mondal@inaf.it}\\
             $^2$Max-Planck-Institut für extraterrestrische Physik, Gießenbachstraße 1, 85748, Garching, Germany\\
             $^3$Columbia Astrophysics Laboratory, Columbia University, New York, NY 10027, USA\\
             $^4$Institute of Space Sciences (ICE, CSIC), Campus UAB, Carrer de Can Magrans s/n, E-08193 Barcelona, Spain\\
             $^5$Institut d’Estudis Espacials de Catalunya (IEEC), Carrer Gran Capità 2–4, E-08034 Barcelona, Spain
             }

   \date{Received XXX; accepted YYY}
   \authorrunning{Mondal et al.}
   \titlerunning{IP in GC}

 
  \abstract
   {For the past decade, it has been suggested that intermediate polars (IPs), a subclass of magnetic cataclysmic variables (CVs), are one of the main contributors to the hard diffuse X-ray emission from the Galactic center (GC) and Galactic ridge.}
   {In our ongoing \xmm survey of the central region of the Galactic disk ($20\degr\times2\degr$), we detected a persistent IP candidate, 1.7\degr\ away from the GC. In this work, we better characterize the behavior of this source by looking at the new and archival \xmm data.}
   {We performed a detailed X-ray spectral modeling of the source. Furthermore, we searched for X-ray pulsations in the light curve as well as its counterpart at other wavelengths.}
   {The \emph{XMM-Newton} spectrum (0.8--10 keV) of the source is described by a partial covering collisionally ionized diffuse gas with plasma temperature $kT=15.7^{+20.9}_{-3.6}$ keV. In addition, the spectrum shows the presence of iron lines at $E=6.44$, 6.65, and 6.92 keV with equivalent widths of $194^{+89}_{-70}$, $115^{+79}_{-75}$, and $98^{+93}_{-74}$ eV, respectively. The X-ray light curve shows a coherent modulation with a period of $P=432.44\pm0.36$ s, which we infer is the spin period of the white dwarf. The white dwarf mass estimated from fitting a physical model to the spectrum results in $M_{\rm WD}=1.05^{+0.16}_{-0.21}\ M_{\odot}$. We were able to find a likely optical counterpart in the \gaia catalog with a G magnitude of 19.26, and the distance to the source derived from the measured \gaia parallax is $\sim$4.3 kpc.}
   {We provide an improved source localization with subarcsec accuracy. The spectral modeling of the source indicates the presence of intervening circumstellar gas, which absorbs the soft X-ray photons. The measured equivalent width of the iron lines and the detection of the spin period in the light curve are consistent with those from IPs.}


   \maketitle
   
%
\section{Introduction}
Intermediate polars (IPs) are a subclass of magnetic cataclysmic variables (CVs) in which the white dwarf (WD) accretes mass from a late-type main-sequence star \citep[see][for review]{copper1990,petterson1994,mukai2017}. The matter from the companion star forms an accretion disk. The accreted matter is funneled onto the WD polar cap region following the magnetic field lines. It releases gravitational energy, creating an intensely bright spot at or above the magnetic poles. The ionized matter follows the magnetic field lines, and as it approaches the WD surface, the radial infall velocity increases to a supersonic speed of 3000--10000 km s$^{-1}$. Therefore, a shock front is developed above the star, and the infalling gas releases its energy in the shock resulting in hard X-ray photons. If the WD spin and the magnetic axis are not aligned, then the star is an oblique rotator and produces the modulation in X-ray emission as it sweeps its lighthouse beam around the sky \citep{watson1985,kim1995}. The X-ray and optical emission from the IPs also show a periodic variation on the timescale of its orbital period around its donor star \citep{zharikov2001,scaringi2011}. So far, around 50 IPs have been identified, and more than 100 candidates are known \citep{downes2001}.

Recently, IPs have been discussed in the context of hard X-ray emission in the Galactic center (GC), bulge, and ridge regions \citep{krivonos2007,revnivtsev2009,hong2012,ponti2013,perez2015,hailey2016}. Early X-ray observations of the GC pointed out the presence of diffuse X-ray emission \citep{yamauchi1996}, so-called Galactic ridge emission with a soft $\sim0.8$ keV and a hard $\sim8$ keV component \citep{komaya1996,tanaka2000}. The soft component is thought to be produced by the large-scale shock waves of supernova explosions, which heat the interstellar medium  \citep{kaneda1997}. The origin of the 8 keV plasma is less certain. On the other hand, the gravitational well of the Galactic disk is too shallow to confine such a hot plasma, which would escape at velocities of thousands of km/s. Therefore, the energy required to sustain it could be equivalent to the release of kinetic energy from one supernova occurring every 30 years \citep{kaneda1997,yamasaki1997}. However, such a rate appears higher than current estimates \citep{ponti2015}. An alternative hypothesis is that a hot plasma is associated with a multitude of faint X-ray sources. The Deep Chandra observations of the central region of the GC ($2\degr\times0.8\degr$; \citealt{wang2002}) and a region south of the GC ($16\arcmin\times16\arcmin$; \citealt{revnivtsev2009}) pointed out that the observed hard ($kT\sim8$ keV) diffuse emission is indeed primarily produced by a population of unresolved magnetic CVs. 

Intermediate polars located at the GC are faint with typical 0.2--10 keV un-absorbed fluxes of $10^{-12}$ erg s$^{-1}$ cm$^{-2}$. The high extinction toward the GC makes it difficult to detect them in soft X-rays. Therefore, deep observations are needed. Currently, we are performing an \xmm survey of the Galactic disk (central $20\degr\times2\degr$). Our \xmm observations have exposures of $\sim$20 ks per tile, which enable us to detect faint point sources. In early 2021, a persistent X-ray point source in the Galactic disk, 1.7\degr\ away from the GC, was detected by \xmm in our ongoing survey. Previously, this field was observed by \swift, \suzaku, and \xmm and the source Suzaku\,J174035.6--301416 has been covered multiple times. \suzaku observed this source twice in 2008 and 2009 \citep{uchiyama2011}. The 2009 observation was not reliable for obtaining the source location as the satellite's altitude was not locked due to a star tracker problem. \cite{uchiyama2011} analyzed the 2008 \suzaku observation; their reported source location is $50.5''$ away from our \xmm position. Considering a 14\arcsec\ positional uncertainty for \suzaku including the statistical and systematic errors \citep{uchiyama2008}, the \suzaku position is off from \xmm at more than a 3$\sigma$ level. These \emph{Suzaku} detections need to be followed up on for better source localization, better statistics, and higher quality spectral and timing data in the X-ray band. \cite{uchiyama2011} performed a detailed spectral and timing analysis of the 2008 \emph{Suzaku} observation. They detected the pulsation of period $432.1\pm0.1$ s and equivalent width of 210, 190, and 130 eV for the 6.4, 6.7, and 6.9 keV lines, respectively. Furthermore, their spectral fitting revealed that a partial covering model is required to fit the spectrum. This paper provides a better source location using the new \xmm observations with improved data quality and reports the results from a detailed spectral and timing modeling.

\section{Observations and data reduction}
This work is primarily based on the 23 ks \xmm observation of the Milky Way plane (ObsID: 0886010601, 2021 March 18). The observation data files were processed using the \xmm \citep{jansen2001} Science Analysis System (SASv19.0.0)\footnote{https://www.cosmos.esa.int/web/xmm-newton/sas}. We used the SAS task \texttt{barycen} to apply the barycenter correction to the event files. We only selected events with PATTERN$\le4$ and PATTERN$\le12$ for EPIC-pn and EPIC-MOS1/MOS2 detectors, respectively. The source and background products were extracted from circular regions with a 25\arcsec\ radius. The background products were extracted from a source-free area. The spectrum from each detector (pn, MOS1, and MOS2) was grouped to have a minimum of 20 counts in each energy bin. Previously, the source had been observed by \xmm on 2018 September 29 (ObsID: 0823030101, 4.1 ks). To gain statistics, we have included the spectra of both observations. The spectral fitting was performed in {\sc xspec} \citep{arnaud1996} and we applied the $\chi^2$ statistic. The spectra from both observations of the EPIC-pn, MOS1, and MOS2 detectors were fitted simultaneously. While fitting the data simultaneously, we added a constant term for cross-calibration uncertainties, which is fixed to unity for EPIC-pn and allowed to vary for MOS1 and MOS2. The best-fit parameters are listed in Table\,\ref{table:fit_tab} with the quoted errors at the 90\% significance level.

\section{Results}
\subsection{X-ray spectra}
\subsubsection{Fitting with phenomenological models}
\label{sec:pheno_models}

\begin{figure}[]
\centering
\includegraphics[width=\figsize\textwidth]{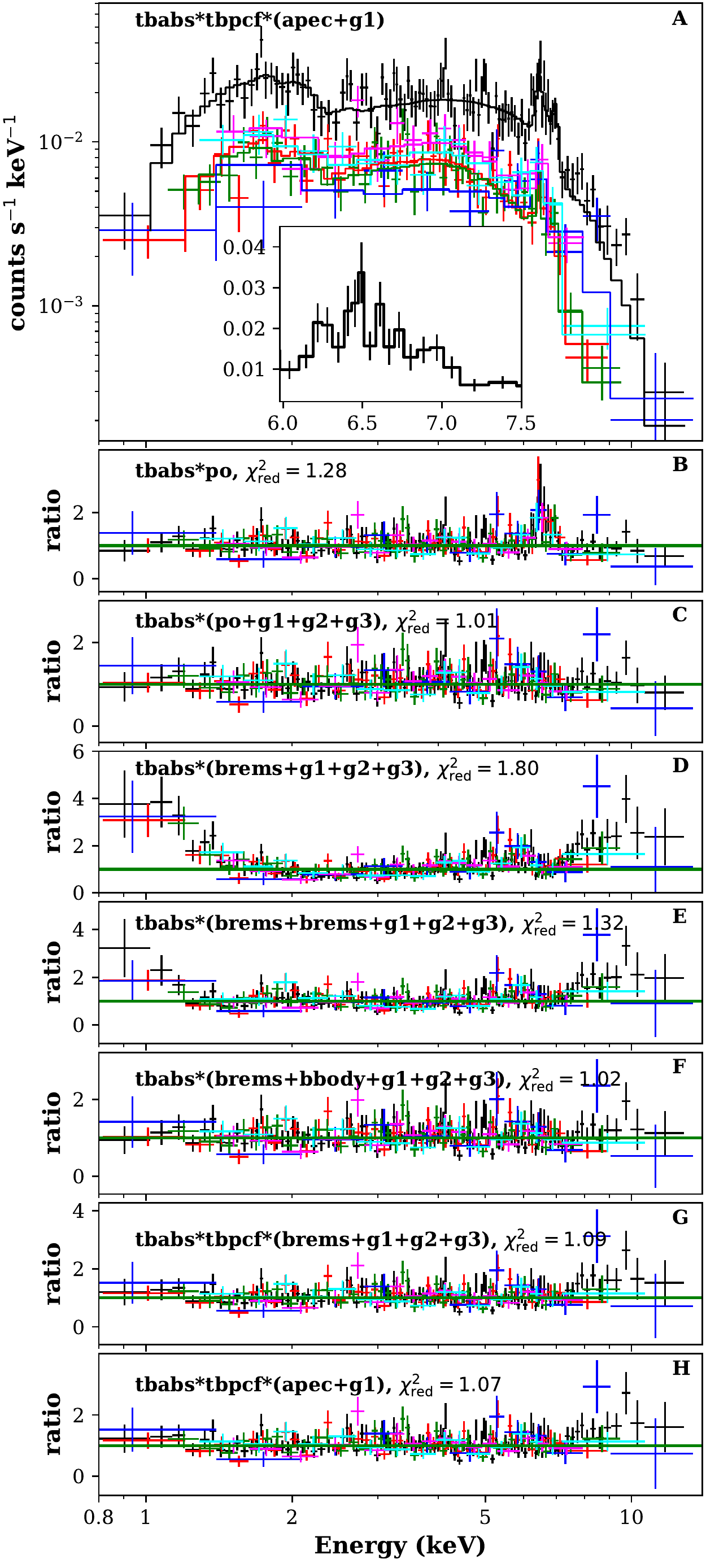}
\caption{Top panel: \xmm time-averaged spectra fitted with a model \texttt{tbabs*tbpcf*(apec+g1)}. The small insert in panel A shows the iron line complex in the 6--7.5 keV region for the EPIC-pn detector of ObsID: 0886010601. The bottom panels show the data-to-model ratio plot obtained from different model fits. The black, red, green, blue, magenta, and cyan data points are from EPIC-pn, MOS1, and MOS2 detectors of ObsID: 0886010601 and 0823030101, respectively.}
\label{fig:ratio}
\end{figure}

First, we fit the spectra with a simple absorbed power-law model (panel B of Fig. \ref{fig:ratio}). A simple power law does not fit the data well $\chi^2_{\rm red}=1.28$. Panel A and B of Fig. \ref{fig:ratio} show that the spectra exhibit multiple iron lines. To fit the iron-line complex, three Gaussian components \texttt{g1}, \texttt{g2}, and \texttt{g3} at 6.44, 6.65, and 6.92 keV, respectively, are required. We froze the width of the Gaussian lines at 0 eV as the current X-ray instruments are unable to resolve the lines. Adding the Gaussian components significantly improved the fit with $\Delta\chi^2=67$ for six additional degrees of freedom ($dof$; panel C of Fig. \ref{fig:ratio}). The simple model \texttt{tbabs*(po+g1+g2+g3)} describes the data quite well with $\chi^2/dof=228/225$ ($\chi^2_{\rm red}=1.01$). Fitting with the power-law model indicates that the source has a very hard spectrum with photon index $\Gamma=0.5\pm0.1$. The hard X-ray spectral index indicates the IP nature of the source. The equivalent widths of the lines at $E=6.44$, 6.65, and 6.92 keV are $194^{+89}_{-70}$, $115^{+79}_{-75}$, and $98^{+93}_{-74}$ eV, respectively. The centroid of the neutral iron $K_{\alpha}$ line is offset from 6.4 keV at the 1$\sigma$ confidence level.

Typically IP spectra are not well described by a power-law model when data beyond 10 keV are available. Therefore, we incorporate more physically motivated models in the following. The spectra from most IPs resemble bremsstrahlung emission with plasma temperatures $\sim$20 keV \citep{ezuka1999}. Therefore, we fit the spectra with the model \texttt{tbabs*(brems+g1+g2+g3)}. The temperature of the bremsstrahlung model is not constrained due to a lack of data above 10 keV. Therefore, we froze the temperature to 20 keV. A simple bremsstrahlung model cannot fit the continuum ($\chi^2/dof=413/229$), leaving an excess below 2 keV and above 7 keV. This is illustrated in panel D of Fig. \ref{fig:ratio}. Adding another bremsstrahlung component significantly improves the fit by $\Delta \chi^2=113$ for two additional $dof$. However, the excess at around 1 keV is still present (see panel E of Fig. \ref{fig:ratio}). The best-fit temperature of the second bremsstrahlung component is much lower than the first one, around $kT=0.17\pm0.02$ keV. This model does not reproduce the data adequately and can be rejected with $P_{\rm Null}=8.8\times10^{-4}$ with $\chi^2/dof=300/227$. The spectrum at higher energies is slightly curved. This curvature cannot be fitted by two bremsstrahlung components absorbed by the same column density of neutral material. We noticed that adding a blackbody component to the bremsstrahlung model gives an excellent fit with $\chi^2/dof=228/224$ (panel F of Fig. \ref{fig:ratio}). However, the blackbody component has a very high temperature of $2.98^{+0.73}_{-0.46}$ keV, as a consequence of trying to reproduce the high energy part of the spectra (>5 keV), which we deem as unlikely. Indeed, a blackbody component is present in some IPs, which have a very soft component in their spectrum. This can be fitted by a blackbody with temperatures of 40--60 eV \citep{haberl1995}, which is not the case here. Next, we added a partial covering absorption component to the bremsstrahlung model to mimic a scenario in which the continuum emission is partially covered by the intervening medium. The model \texttt{tbabs*tbpcf*(brems+g1+g1+g3)}  provides an acceptable fit with $\chi^2/dof=245/224$ ($\chi^2_{\rm red}=1.09$; panel F of Fig. \ref{fig:ratio}). The $N_{\rm H,pcf}$ value is almost an order higher than the Galactic absorption. To constrain the plasma temperature, we fit the data with a partially covered \texttt{apec} model plus a Gaussian line at 6.4 keV. This model may provide a better estimate of the plasma temperature by using both the continuum as well as the 6.7 and 6.9 keV line ratio. This model provides a better fit than the bremsstrahlung model with $\chi^2/dof=244/227\ (\chi^2_{\rm red} = 1.07$; panel H) and the plasma temperature obtained from this model is $kT=15.65^{+20.87}_{-3.58}$ keV.

\begin{figure}[]
\centering
\includegraphics[width=\figsize\textwidth]{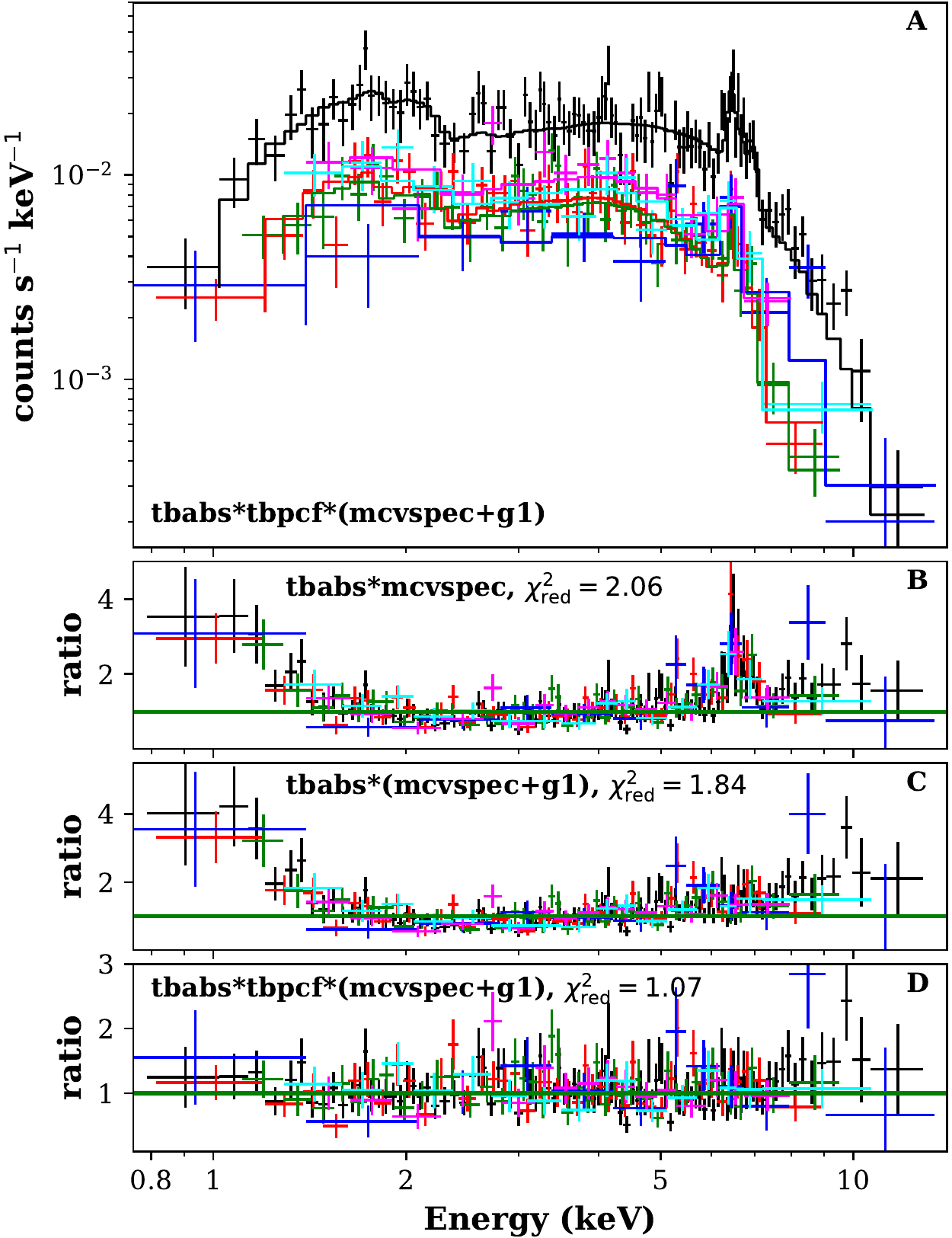}
\caption{Same as Fig. \ref{fig:ratio}, but for the model \texttt{tbabs*tbpcf*( mcvspec+g1)}.}
\label{fig:mcvspec_spec}
\end{figure}

\subsubsection{Fitting with the physical model \texttt{mcvspec}}
We also fit the data with the \texttt{mcvspec} model. This is a physical model which reproduces the emission from the surface of a magnetized WD. The model is available in {\sc xspec} and is based on the one-dimensional accretion column model of \citealt{saxton2005} (Mori et al. in preparation). The model includes lines produced by collisionally ionized diffuse gas in the accretion column of the WD. While doing the fitting, we noticed that there is a degeneracy between magnetic field $B$ and the mass accretion flux $\dot m$. Therefore, we froze the $B$ and $\dot m$ to values 10 MG and 5 g cm$^{-2}$ s$^{-1}$, respectively, which is typically observed in IPs (Mori et al.). The \texttt{mcvspec} model includes the line from ionized Fe; however, it does not include the neutral 6.4 keV Fe $K_{\alpha}$ line. Therefore, we added a Gaussian at 6.4 keV with the width of the line frozen to zero. In the fitting procedure, the abundance value was frozen at the solar value.

Initially, we tried to fit the data with this model with a single absorption component. As we saw for the bremsstrahlung and \texttt{apec} models,   the partial covering of the source is also required for
the \texttt{mcvspec} model. The neutral 6.4 keV line is also clearly visible in panels A and B of Fig. \ref{fig:mcvspec_spec}. When we added a Gaussian component at 6.4, the excess between 6--7 keV was resolved (shown in panel C of Fig. \ref{fig:mcvspec_spec}). The spectral fit is shown in Fig. \ref{fig:mcvspec_spec}. The WD mass estimated from fitting this model is $1.05^{+0.16}_{-0.21}\ M_{\odot}$.

\begin{table}
\caption{Best-fit parameters of the fitted models.}
\label{table:fit_tab}
\setlength{\tabcolsep}{1.1pt}                   
\renewcommand{\arraystretch}{1.4}               
\centering
\begin{tabular}{c c | c c}
\hline\hline
 \multicolumn{2}{c|}{\multirow{2}{*}{\texttt{tbabs*tbpcf*(apec+g1)}}} & \multicolumn{2}{c}{\texttt{tbabs*tbpcf*(mcvspec+}}\\
 && \multicolumn{2}{c}{\texttt{g1)}}\\
 \hline
 $N_{\rm H}$ & $1.29^{+0.19}_{-0.16}$ & $N_{\rm H}$ & $1.30^{+0.20}_{-0.16}$\\ 
 $pcf$ & $0.69^{+0.04}_{-0.06}$ & $pcf$ & $0.67^{+0.05}_{-0.04}$\\
 $N_{\rm H,pcf}$ & $15.13^{+8.42}_{-3.68}$ & $N_{\rm H,pcf}$ & $15.82^{+7.15}_{-4.17}$\\
 $kT$ & $15.65^{+20.87}_{-3.58}$ & $B$ & $10\times10^6$ (frozen)\\
 Abun & 1.0 (frozen) & Abun & 1.0 (frozen)\\
 Redshift & 0.0 (frozen) & $\dot m$ & 5 (frozen)\\
 $N_{\rm apec}$ & $1.94^{+0.30}_{-0.19}\times10^{-3}$ & $M$ & $1.05^{+0.16}_{-0.21}$\\
 $E_{\rm g1}$ & $6.44^{+0.05}_{-0.03}$ & $N_{\rm mcv}$ & $1.58^{+0.71}_{-0.20}\times10^4$\\
 $N_{\rm g1}$ & $9.81^{+2.38}_{-2.37}\times10^{-6}$ & $E_{\rm g1}$ & $6.44^{+0.04}_{-0.03}$\\
 && $N_{\rm g1}$ & $7.68^{+1.90}_{-1.02}\times10^{-6}$\\
 
 $F_{\rm x}$ & $1.92^{+0.15}_{-0.15}\times10^{-12}$ & $F_{\rm x}$ & $1.90^{+0.14}_{-0.14}\times10^{-12}$\\
 
 $\chi^2_{\rm red}\ (dof)$ & 1.07 (227) & $\chi^2_{\rm red}$ & 1.07 (227) \\
\hline
\end{tabular}
\tablefoot{The $N_{\rm H}$ is given in units of $10^{22}$ cm$^{-2}$. The metal abundance value is frozen to 1.0. Due to degeneracy, we froze the $B$ and $\dot m$ to 10 MG and 5 g cm$^{-2}$ s$^{-1}$, respectively. The observed X-ray flux $F_{\rm x}$ is given in the 0.8--10 keV band.}
\end{table}

\subsection{Periodicity search}
\begin{figure}[]
\centering
\includegraphics[width=\figsize\textwidth]{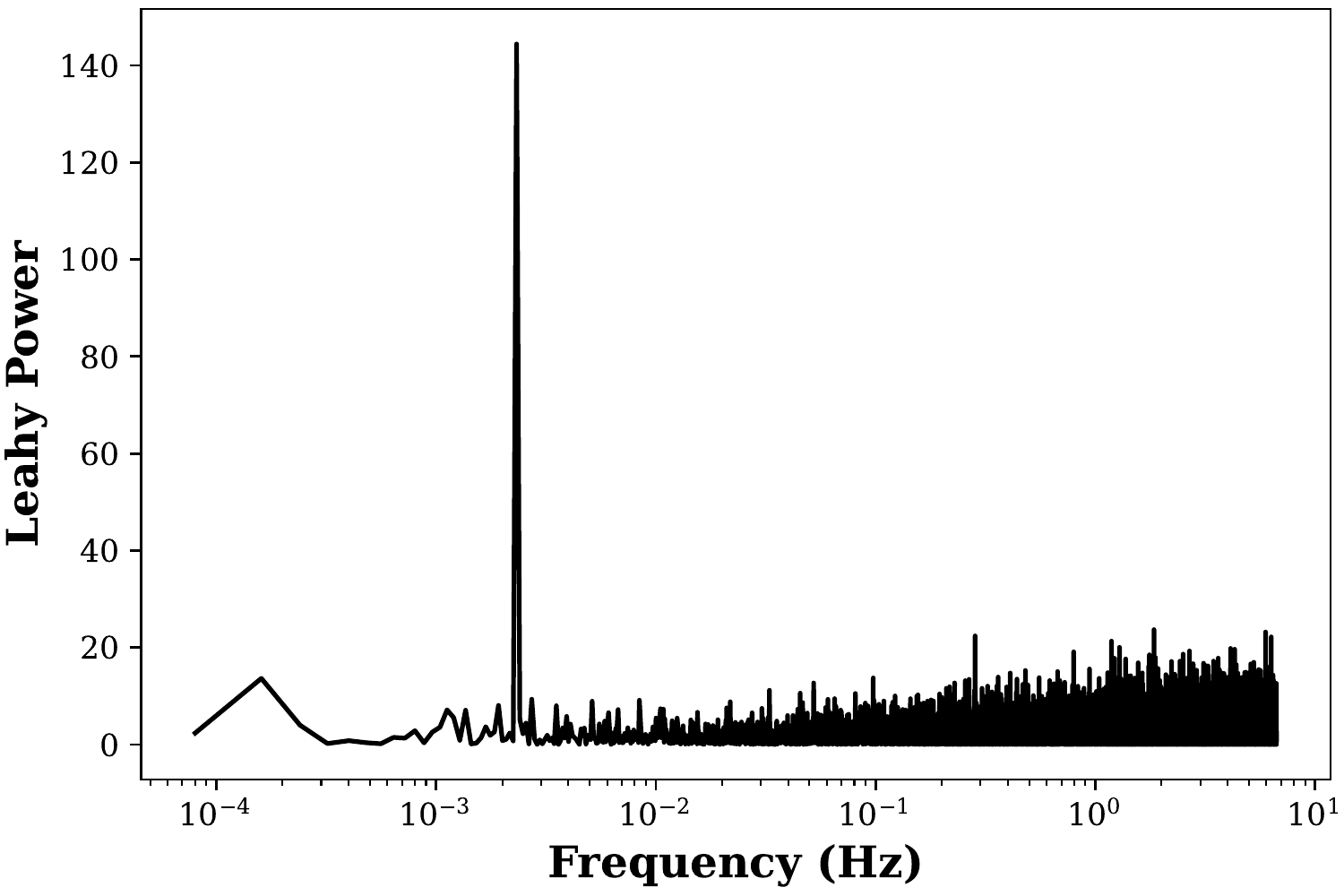}
\caption{Fast Fourier transform computed from the longest continuous good time interval (GTI) of the EPIC-pn data. The PSD shows a clear peak at $2.3\times10^{-3}$ Hz, which is likely associated with the spin frequency of the WD.}
\label{fig:PSD}
\end{figure}

\begin{figure}[]
\centering
\includegraphics[width=\figsize\textwidth]{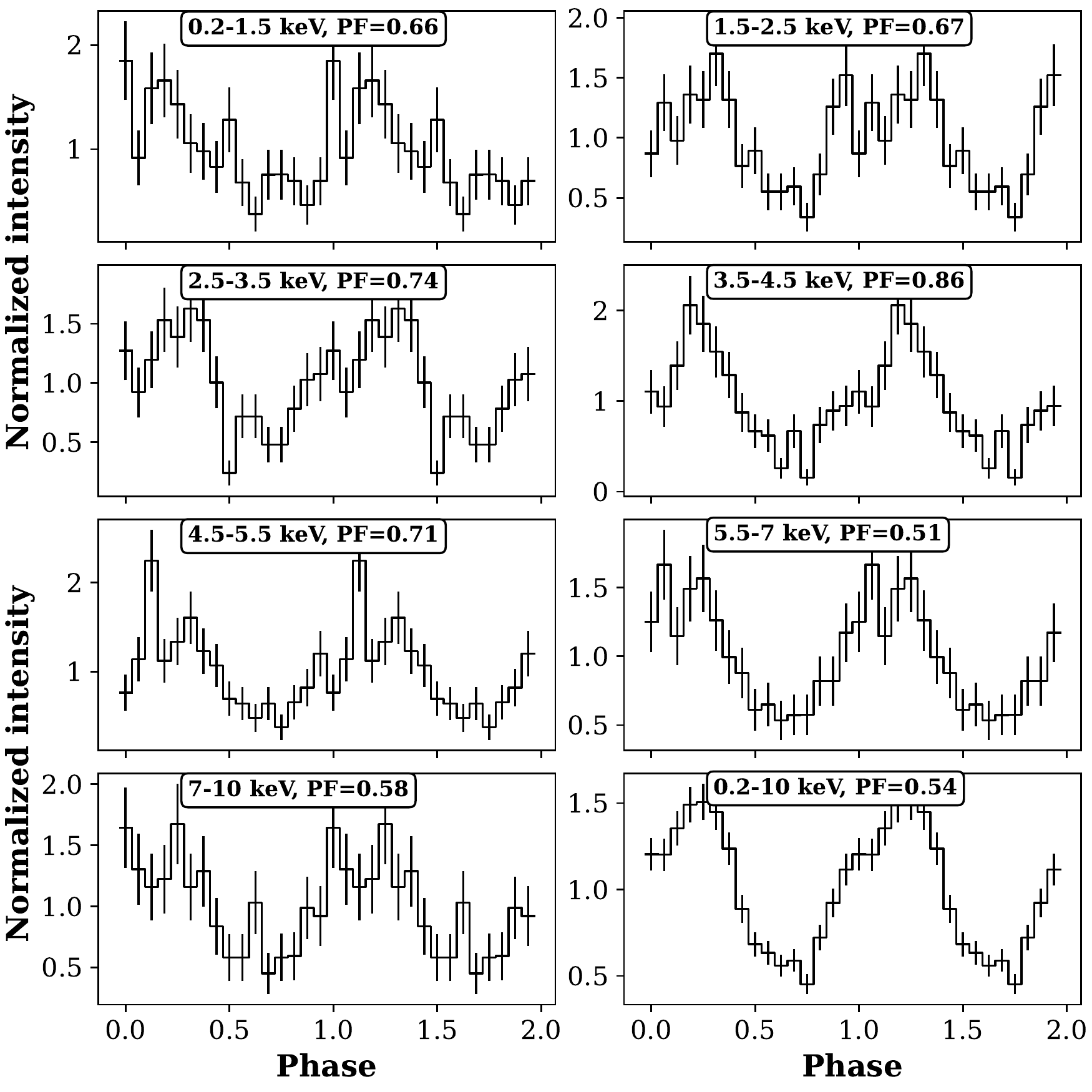}
\caption{Complex shape of the pulse profile at different energy bands. The light curves were folded with a period of 432.44 s.}
\label{fig:phase}
\end{figure}

For our timing analysis, we focused on the data from the EPIC-pn detector only as it has the shortest frame time, which allows us to examine the highest possible frequency range. We extracted the light curve with a time bin of 74 ms. We computed the fast Fourier transform from the longest continuous segment length of 12.5 ks; with this setup, we could probe frequencies from $8.1\times10^{-5}$ Hz to 13.5 Hz. A narrow peak at a frequency of  $2.3\times10^{-3}$ Hz is clearly visible in the power spectral density (PSD) (Fig.\,\ref{fig:PSD}). We applied the Bayesian formalism of \citet{1992ApJ...398..146G} to refine the period further and estimate the error. This method is optimized for the detection of a periodic signal by comparing different periodic models for a given frequency and phase bins with a model of a constant count rate. We selected energies between 0.2 keV and 8 keV and set the maximum number of phase bins to ten. This resulted in the most probable period of 432.44$\pm$0.36 s. We folded the light curves in different energy bands with the period derived from the Bayesian method. The pulse profiles at various energy bands are shown in Fig. \ref{fig:phase}. The energy dependence of the pulse profile is complex. The pulse profiles at lower energies appear to be nonsinusoidal and nearly sinusoidal at higher energies. Additionally, the pulsed fraction ($\rm PF=\frac{F_{\rm max}-F_{\rm min}}{F_{\rm max}+F_{\rm min}}$) is higher at energies below 5.5 keV.

\subsection{The long-term X-ray flux variability and multiwavelength counterpart}
\begin{figure}[]
\centering
\includegraphics[width=\figsize\textwidth]{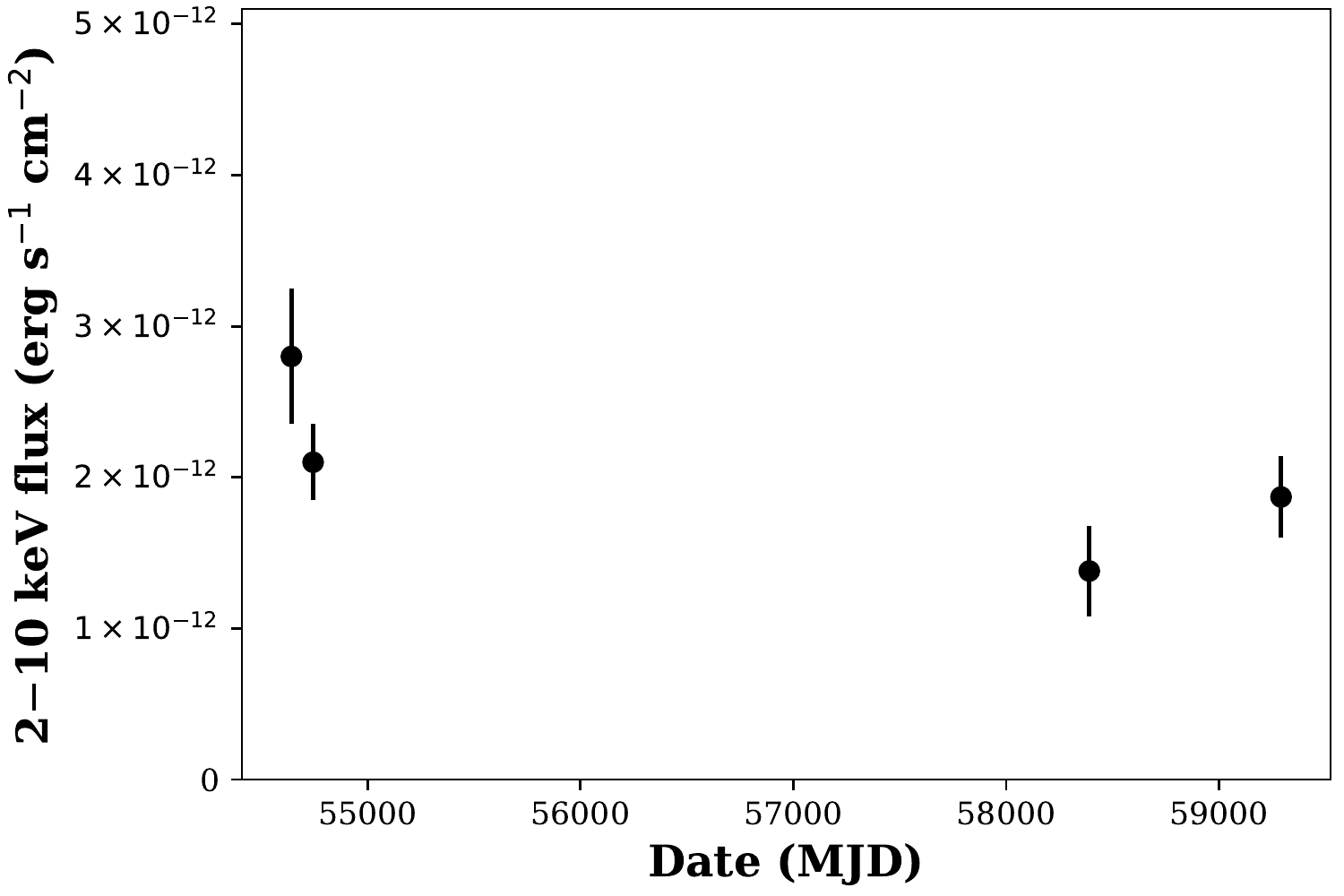}
\caption{Long-term light curve of the source. The 2--10 keV flux given here is the observed flux. The source does not exhibit significant flux variability.}
\label{fig:long_lc}
\end{figure}

The source is located 1.71\degr\ away from the Galactic center. The source position after the bore-sight correction is $(\alpha,\delta)_{\rm J200}$ = (17\rahour\,40\ramin\,33\fs85, --30\degr\,15\arcmin\,01\arcsec) which is designated as 4XMM\,J174033.8--301501 with a 1$\sigma$ positional uncertainty of $0.5''$. A counterpart in \swift-XRT was also detected in 2008 June 25 observations. A bright source with an offset of 4\arcsec\ from the \xmm position is listed in the \swift-XRT point source catalog \citep{delia2013}. Fig. \ref{fig:long_lc} shows the long-term light curve of the source. The \swift-XRT 2--10 keV observed flux is $2.8\times10^{-12}$ erg s$^{-1}$. \suzaku also observed this source on 2008 October 05 and 2009 February 26. \cite{uchiyama2011} reported that the \suzaku 2--10 keV flux in the 2008 observation is around $\sim2\times10^{-12}$ erg s$^{-1}$ cm$^{-2}$. The 2--10 keV flux in 2018 and the 2021 \xmm observations are $1.38\times10^{-12}$ erg s$^{-1}$ cm$^{-2}$ and $1.87\times10^{-12}$ erg s$^{-1}$ cm$^{-2}$, respectively. This indicates that even though the source has been detected multiple times spanning more than a decade, it shows very little to no flux variability. 

We searched for counterparts at other wavelengths.  Our \xmm observation has simultaneous UV observation with filters UVW1 (291 nm), UVW2 (212 nm), and UVM2 (231 nm). We did not find any counterparts in our UV observation within a 3$\sigma$ positional uncertainty (1.5\arcsec) of the EPIC detection. The 2008 \swift observation was done with UVM2 filter (224.6 nm). No counterpart was found in the \swift UVM2 observation either. Furthermore, we searched for counterparts in the \gaia and 2MASS catalog. A likely associated \gaia counterpart is 0.48\arcsec\ away with a G magnitude of $19.262\pm0.005$. The parallax and proper motion obtained from \gaia are 0.23 milli-arcsec and 1.92 milli-arcsec yr$^{-1}$, respectively. Given the measured parallax by \gaia, the distance to the source is 4.3 kpc. No counterpart was found in the infrared 2MASS catalog.

\section{Discussion}
 The spectrum of the source is complex and shows the presence of iron lines at energies 6.44, 6.65, and 6.92 keV with an equivalent width of $194^{+89}_{-70}$, $115^{+79}_{-75}$, and $98^{+93}_{-74}$ eV, respectively. The measured equivalent width of the lines is similar to the values found in magnetic CVs \citep{hellier2004}. We also detected a clear, coherent pulsation with a pulse period of 432.44 s. The detected spin period is well within the typical range for magnetic CVs, which is 30 s to $2\times10^{4}$ s \citep{scaringi2010}. The equivalent width and spin period based on the \suzaku observation are consistent with our measurements. The spin period, iron line complex, and the equivalent width of the iron lines indicate the source is a strong candidate for intermediate polar. The source is less likely to be polar, as polars have a very high magnetic field of 10--240 MG, which synchronizes the spin and orbital period. The measured 432.44 s period is too short to be the synchronized spin-orbital period; in polars, the spin-orbital period typically ranges from (0.3--6)$\times10^{4}$ s \citep{scaringi2010}. For intermediate polars, the spin-to-orbital period ratio is around 0.1 \citep{barrett1988}; therefore, in this case, the orbital period would be on the order of 4500 s. However, we do not see a large peak at the corresponding frequency in the PSD, but this might be because the longest continuous GTI only covers two cycles for the orbital period. 

Two absorption-component models are required to fit the overall spectrum of the source. This is also noted by \cite{uchiyama2011} who, while fitting \suzaku data, required two absorption components to mimic the partial covering scenario. \cite{uchiyama2011} first tried to fit the data with a single \texttt{apec} model, which indicates a plasma temperature of $9.5^{+1.5}_{-0.2}$ keV. The authors further noticed that adding another \texttt{apec} component significantly improves the fit. For the two-temperature model, the temperature of the two \texttt{apec} components are $6\pm1$ keV and $kT\sim64$ keV. Our spectral fitting results are fully consistent with the findings of \cite{uchiyama2011}. Our best-fit \texttt{apec} model gives plasma temperature $kT=15.13^{+20.87}_{-3.58}$ keV, which is similar to the value found by \cite{uchiyama2011}. The centroid of the neutral iron $K_{\alpha}$ line is $6.44^{+0.05}_{-0.03}$, which is shifted from 6.4 keV at a $1\sigma$ confidence level. The shift is at the level of 1$\sigma$. This shift can be explained by assuming that the material is not completely cold and is slightly ionized with $\xi\sim$ 200--300 \citep{matt1993}. We tested a physically motivated model to constrain the mass of the central compact object. The mass of the central compact object obtained from this model is $1.05^{+0.16}_{-0.21}\ M_{\odot}$. It is pretty clear that no matter which physical model we use, a partial covering with high absorption column density $N_{\rm H}\sim1.5\times10^{23}$ cm$^{-2}$ is required to fit the continuum. We suggest this large amount of absorption is due to circumstellar gas.

From an evolutionary point of view, IP will have a cool main-sequence companion star, which should be identified as a blue optical counterpart \citep{hong2009}. Estimating the X-ray-to-optical flux ratio using \gaia G band magnitude is possible. We used the formula ${\rm Log}(F_{\rm x}/F_{\rm G})={\rm Log}(F_{\rm x})+m_{\rm G}/2.5+5.37$ \citep{maccacaro1988}, where $F_{\rm x}=1.87\times10^{-12}\ \rm erg\ cm^{-2}\ s^{-1}$ is 2--10 keV flux and $m_{\rm G}=19.26$ is the \gaia G band magnitude. This gives ${\rm Log}(F_{\rm x}/F_{\rm G})=1.35$ which is consistent with the values found in Galactic bulge CVs \citep{bahramian2021}. Previously, other X-ray satellites have detected the source many times; however, the source shows no flux variability. In general, IPs display some level of long-term variability in both the X-ray and optical. The expected luminosity range of IPs are in between $3\times10^{29}$--$5\times10^{33}$ erg s$^{-1}$ \citep{ruiter2006}. The 0.8--10 keV observed luminosity of the source is $4.05\times10^{33}$ erg s$^{-1}$, which places it in the brightest class of IPs. The low-luminosity IPs are difficult to detect due to the large absorption toward the GC. The measured X-ray interstellar medium absorption $1.3\times10^{22}$ cm$^{-2}$ is consistent with the expected total column density of the Galaxy ($N_{\rm H} = N_{\rm HI}+N_{\rm H2} = 1.37\times10^{22}$ cm$^{-2}$ , \citealt{willingale2013}) toward the source position. A better characterization of the population of IPs toward the GC can help us to solve the mystery of Galactic diffuse hard X-ray emission.

\section{Conclusions}
In this work, we have provided a better positional accuracy of the source and identified an optical counterpart in the \gaia catalog. It is a source located at a distance of 4.2 kpc. The spin period of the WD is $432.44\pm0.36$ s, and the orbital period was not recovered from the X-ray light curve. Previously, the source had been observed on other occasions, but it did not show any X-ray flux variability. A partial covering collisionally ionized diffuse gas with plasma temperature $15.13^{+20.87}_{-3.58}$ keV fits the data well. We estimated the WD mass to be $1.05^{+0.16}_{-0.21}\ M_{\odot}$. We did not find any counterpart in the UV detector of \xmm and \swift, and also, no IR counterpart was found in the 2MASS catalog.


\begin{acknowledgements}
SM, GP, and KA acknowledge financial support from the European Research Council (ERC) under the European Union’s Horizon 2020 research and innovation program HotMilk (grant agreement No. 865637).

\end{acknowledgements}

%
%

\bibliographystyle{aa} 
\bibliography{refs}

\end{document}